\documentclass[twocolumn]{aastex63}

\usepackage{amsmath,amstext}
\usepackage{mathtools}
\usepackage{afterpage}
\usepackage{lipsum}
\usepackage{relsize}
\usepackage[all]{hypcap} 
\usepackage{verbatim}
\usepackage{lineno}
\usepackage{graphicx}
\usepackage{footnote}
\usepackage{float}
\usepackage{amsmath}
\usepackage{enumitem}
\usepackage{amssymb}
\usepackage{array}
\usepackage{gensymb}
\usepackage{mathtools}
\usepackage{relsize}
\usepackage{lipsum}
\usepackage{lineno}
\usepackage{mathptmx}
\usepackage[graphicx]{realboxes}
\usepackage{natbib}

\usepackage{lipsum}
\usepackage{rotating}
\usepackage{nicefrac}
\newcommand{\CHE}{CHE Israel Excellence Fellowship}

\usepackage{color, colortbl}

\begin{document}

\title{Conserving Local Magnetic Helicity in Numerical Simulations.}
\correspondingauthor{Yossef Zenati}
\email{yzenati1@jhu.edu}

\author[0000-0002-0632-8897]{Yossef Zenati}
\altaffiliation{\CHE}
\affil{Physics and Astronomy Department, Johns Hopkins University, Baltimore, MD 21218, USA}

\author{Ethan T. Vishniac}
\affil{Physics and Astronomy Department, Johns Hopkins University, Baltimore, MD 21218, USA}

\date{April 2021}

\begin{abstract}
Magnetic helicity is robustly conserved in systems with very large magnetic Reynolds numbers, including most systems of astrophysical interest, and unlike kinetic and magnetic energy is not dissipated at small scales. This plays a major role in suppressing the kinematic large-scale dynamo and may also be responsible for driving the large-scale dynamo through the magnetic helicity flux. Numerical simulations of astrophysical systems typically lack sufficient resolution to enforce global magnetic helicity over several dynamical times. In these simulations, magnetic helicity is lost either through numerical errors or through the action of an unrealistically large resistivity. Errors in the internal distribution of magnetic helicity are equally important and typically larger.  Here we propose an algorithm for enforcing strict local conservation of magnetic helicity in the Coulomb gauge in numerical simulations so that their evolution more closely approximates that of real systems.
\end{abstract}

\section{Introduction}
Magnetic helicity is a conserved quantity in ideal MHD \citep{Woltjer58} and one of the primary ways to quantify the complexity of a magnetic field. In a weakly nonideal turbulent system, it is still almost conserved, in the sense that the turbulent cascade is ineffective at transferring it to resistive scales, even while the magnetic and kinetic energies are dissipated efficiently. Consequently, \citet{Taylor74} offered the conjecture that a laboratory system would first relax to the minimum energy state with the same total magnetic helicity (see also \cite{Taylor86}). For example, see \citet{Yamada99} for a review of the laboratory evidence accumulated in the following two decades. We note in particular \citet{Ji96} which shows that the dynamical transport of magnetic helicity is an important part of this balance. The conservation of magnetic helicity
should be even more robust in astrophysical systems where the microscopic resistivity is usually negligible on large eddy scales.
Minimizing energy for the same global magnetic helicity drives the creation of large-scale magnetic fields. We can see this in simple terms by noting that the magnetic helicity density has the dimensions of $[length]\times [energy density]= Wb^2$. For large-length scales, the same magnetic helicity requires the least energy. This same argument shows why magnetic helicity is so resistant to dissipation. It can't participate in the turbulent cascade since the energy density corresponding to small-scale eddies decreases with decreasing eddy size. The tendency of robust magnetic helicity conservation to result in an inverse cascade motivates the proposal that the large-scale dynamo ($\rm LSD$) is driven by magnetic helicity \citep{Vishniac_Cho01}
though a combination of turbulent transport and the inverse cascade. 
One of the main challenges in testing this proposal is to examine the flow of magnetic helicity spatially and between scales when numerical simulations tend to dissipate magnetic helicity at significant rates.

In addition to its importance in determining the relaxation of complex systems, magnetic helicity may play a role in the large-scale dynamo process, a point that was first raised by \citet{Pouquet76}. The evolution of the large scale magnetic field ${\bf B_l}$, where the subscript denotes low pass filtering on the scale $l$, is given by
\begin{equation}
    \partial_t{\bf B_l}={\bf \nabla\times(\bf v_l\times B_l-\eta{\bf J_l})}+{\bf\nabla\times e_l},
\end{equation}

where
\begin{equation}
    {\bf e_l}\equiv ({\bf v\times B})_l-{\bf v_l\times B_l}.
\end{equation}

And Ohm’s law is
\begin{equation}
    {\bf E}\equiv {\bf -v_{l}\times B_{l}}+{\eta \bf J_l}.
\end{equation}

$\eta$ is the resistivity for a magnetic diffusivity. In a turbulent medium ${\bf e_l}$, the electromotive force filtered on the scale $l$, is approximately the large-scale average of the cross product of the turbulent velocity and magnetic fields. \citet{Pouquet76} pointed out that ${\bf e_l}$ has contributions from the eddy scale contributions to the averaged kinetic and current helicities multiplied by ${\bf B_l}$, i.e. as a part of the $\alpha$ effect in dynamo theory. Neither helicity is conserved in ideal MHD but in the Coulomb gauge the eddy scale current helicity, ${\bf j\cdot b}$, is approximately $k^2$ times the eddy scale magnetic helicity, ${\bf a\cdot b}$, where $k^2$ is the mean square wavenumber associated with the large scale eddies. While only the total magnetic helicity is conserved, the transfer of magnetic helicity between large and small scales depends on the parallel component of the electromotive force. This connection was exploited by \citet{Gruzinov96} to show that a dynamo driven by the kinetic helicity saturates due to the local accumulation of magnetic helicity, and turns off long before the large scale field reaches equipartition with the turbulent energy density ("$\alpha$ suppression"). Apparently, the conservation of magnetic helicity can have profound implications for the large-scale dynamo.

Given the existence of large-scale magnetic fields, it seems safe to assume that there is some way to avoid the negative conclusion offered by \citet{Gruzinov96}. Possibly the kinematic dynamo operates, perhaps at a reduced efficiency, while the eddy scale magnetic helicity is ejected from the system (for a discussion of this idea see \citet{brandenburg09a}). Alternatively, the kinetic helicity may be irrelevant, and a magnetic helicity flux, induced by rotation, creates local accumulations of eddy scale magnetic helicity which then drive a large scale dynamo as a side effect of being transferred to large scale fields \citep{Vishniac_Cho01}. Finally, it is possible that a large-scale dynamo ($\rm LSD$) could be produced through a combination of shear and fluctuating kinetic helicity \citep{Vishniac_Brandengurg97, Mitra_Brandenburg12}, although the relation between this mechanism and $\alpha$ suppression has never been fully explored.

Regardless of which of these mechanisms dominate in real systems, we can see that accurate simulations of dynamos require careful accounting of the distribution of magnetic helicity in a system. To model Taylor relaxation accurately one needs only conserve the total magnetic helicity in a simulation box. To model dynamos, including $\alpha$ suppression and the transport of magnetic helicity, one needs to extend this down to scales not much larger than individual eddy scales. Moreover, since the magnetic helicity is a quadratic quantity, the average magnetic helicity on domain scales can arise from correlations on eddy scales. The dissipation of this contribution to the magnetic helicity proceeds at the ohmic dissipation time at the large eddy scale (or a bit faster, see below). Numerical dissipation on this scale has to be slower than the large-scale dynamo growth rate for an accurate large-scale dynamo simulation. In practice, while this is a trivial threshold for real systems, it represents a substantial challenge for simulations. This is a particularly difficult issue for simulations of magnetorotational instability, where the large-scale eddy size is proportional to the strength of the magnetic field. Worse is that the MRI operates primarily in the $\hat r \hat \phi$ plane, and the vertical wave number in unstratified simulations tend to the dissipation scale, which guarantees that magnetic helicity will not be conserved for more than an eddy correlation time unless the magnetic Prandtl number is larger than one \citep[see,][]{Brandenburg+05}.

This raises the question of whether it might be useful to enforce magnetic helicity conservation in numerical simulations. In many cases, it is possible to enforce the conservation of an important conserved quantity by choosing an appropriate numerical algorithm, but magnetic helicity is nonlocal and no local numerical algorithm conserves it exactly. Evaluating magnetic helicity in simulations of the MRI in a stratified accretion disk, for example, shows that in a single orbit, the total magnetic helicity in the box can change by a number comparable to a density scale height times the magnetic energy density \citep[see,][]{Davis+10}.
What price do we pay for adjusting the small-scale magnetic field in real-time to compensate for numerical errors? As a general rule, requiring strict conservation of any quantity will prioritize it above other conserved quantities. In particular, adjusting the magnetic field to conserve magnetic helicity will affect the stress-energy tensor, including the energy density. However, in a turbulent simulation, the fluctuating magnetic energy is continuously lost to the turbulent cascade, so energy conservation at small length scales is of secondary importance unless the energy input (or loss) is comparable to the turbulent energy dissipation rate. Any scheme that adjusts local magnetic field strengths will violate flux freezing, but this is already violated by turbulent transport in the turbulent cascade \citep{eyink13}. Lastly, if there is no turbulence, then there's almost no numerical loss of magnetic helicity and any scheme for correcting the field will have a negligible effect. In principle, there seems to be no fundamental obstacle to implementing a correction scheme to enforce magnetic helicity conservation. Whether or not a particular scheme improves the accuracy of numerical simulations at an acceptable cost depends on the details of the algorithm. In this paper, we will suggest an approach to enforcing magnetic helicity conservation and explore some of the constraints on its use.

In section 2 of this paper, we will review the definition of magnetic helicity, and other relevant quantities, and their evolution in a turbulent medium.  We will review how fluctuations in the magnetic helicity are generated and dissipated, including resistive losses. Finally, we will present an algorithm that allows us to impose strict conservation of magnetic helicity with minimal effect on other properties of the magnetic field. In section 3 we will test the convergence of our algorithm. In section 4 we discuss the applications and limitations of this tool.

\section{Defining an Algorithm}

In a highly conducting fluid, the magnetic field evolves as
\begin{equation}
    \partial_t{\bf B}=\nabla\times\left({\bf v\times B-\eta J}\right),
\end{equation}
where the resistive term $\eta{\bf J}$ is small. Numerical effects may or may not mimic physical resistivity. We will assume here that they do, but note places where the difference might be important. Consequently, the vector potential evolves as
\begin{equation}
    \partial_t{\bf A}={\bf v\times B-\eta J}-\nabla\phi,  
\end{equation}
where the electrostatic potential,$\phi$ satisfies
\begin{equation}
    \nabla^2\phi=\nabla\cdot(\bf{v\times B-\eta J}),
\end{equation}
in the coulomb gauge. This gauge choice has the advantage of enforcing a tight correlation between the small-scale contribution to the total current helicity, ${\bf J\cdot B}$, and the small-scale contribution to the magnetic helicity. The former is gauge invariant and plays an important role in dynamo theory. We note that other gauge choices have been used for computational convenience or because the magnetic helicity has some compelling interpretation in that gauge (for examples see \citep{Alexander&Axel11, Candelaresi+11, DelSordo+13}).

In this gauge the magnetic helicity flux is
\begin{equation}
{\bf J}_H\equiv {\bf A \times}({\bf v\times B}-\eta J+\nabla\phi),
\label{Hflux}
\end{equation}
and the magnetic helicity, $H$, evolves as
\begin{equation}
    \partial_t H=-\nabla{\bf \cdot J}_H-2\eta{\bf J\cdot B}.
    \label{Hevolve}
\end{equation}
When the resistivity is small its contribution to ${\bf J}_H$ will be negligible, but its role in breaking magnetic helicity conservation can still be important. We can define the small-scale contribution to the magnetic helicity as $h_l\equiv H_l-{\bf A_l\cdot B_l}$ so that the evolution equation is
\begin{equation}
    \partial_t h_l=-\nabla\cdot{\bf J}_{Hl}-2\eta([({\bf J\cdot B})_l-{\bf J}_l\cdot{\bf B}_l] -2{\bf B_l\cdot e_l},
    \label{heltran}
\end{equation}
where the eddy scale magnetic helicity flux, $J_{Hl}$, is defined, as for the other eddy scale contributions, by subtracting from the total flux the contribution from large scale fields, including $e_l$. The last term on the RHS of equation (\ref{heltran}) describes the transfer of magnetic helicity between small and large scales and is critical for understanding how magnetic helicity is dissipated when the resistivity is not exactly zero. 

\begin{figure}
\includegraphics[width=\linewidth]{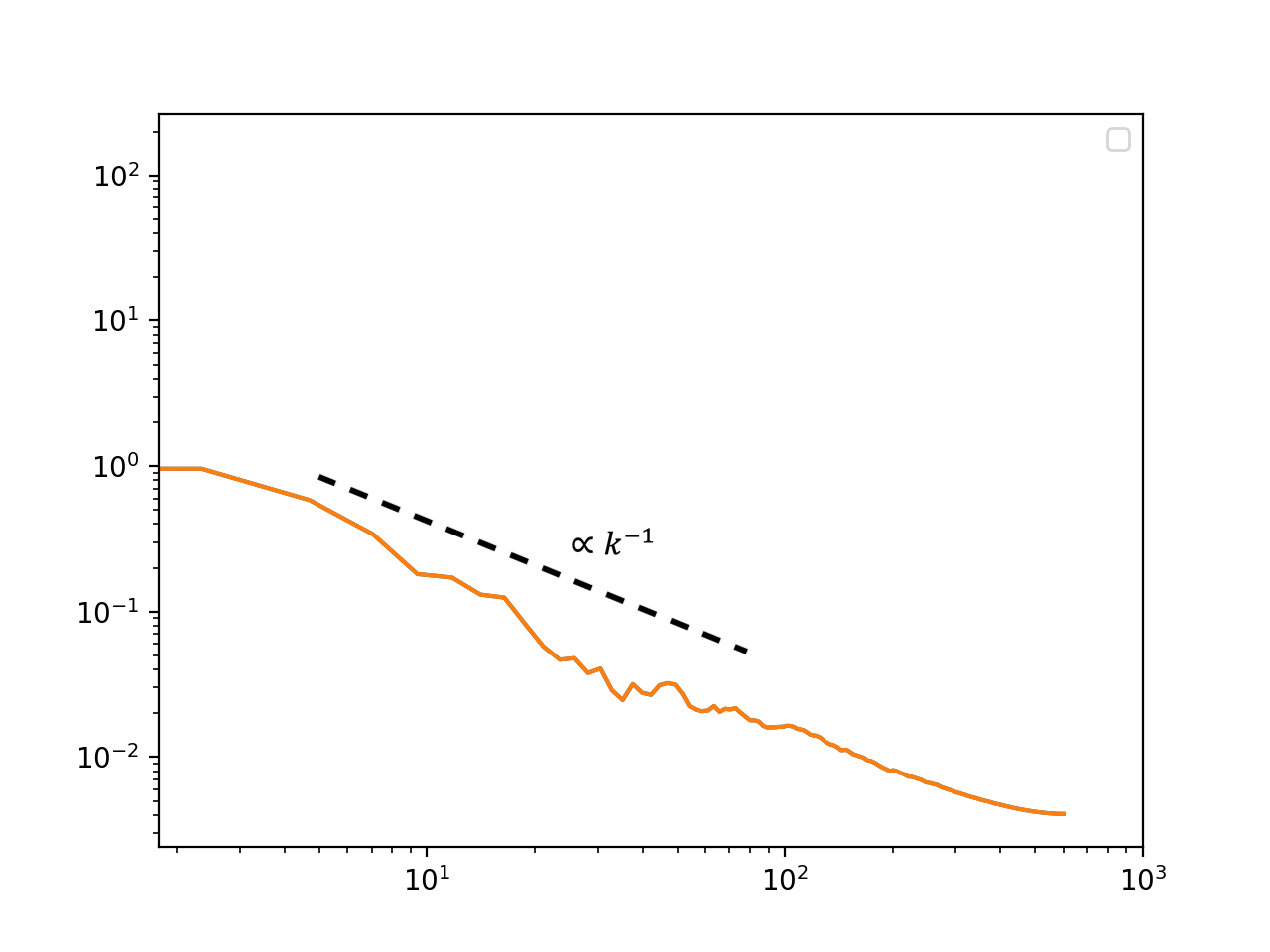}
\caption{The current helicity $H_J$ spectrum compared to the expected $k^{-1}$ one dimensional spectrum.}
\label{fig:kpowerl}
\end{figure}

For this purpose eq.(\ref{heltran}) can be generalized by dividing the turbulent cascade into a series of shells in phase space, each with a typical wave number amplitude $k$ \citep{Aluie2017}. Then instead of large and small scale contributions to the total helicity on the scale $l$ we have contributions from each shell, $h_k$. The transfer of magnetic helicity between large and small scales is still ${\bf B\cdot e_l}$, but now ${\bf e_l}$ is the sum of multiple terms, each given by $({\bf v\times b})_{kl}$ the contribution to ${\bf e_l}$ from the shell $k$ and each term gives the transfer of averaged helicity, $H_l$, between large scales and the scale given by $k^{-1}$ see figure \ref{fig:kpowerl}.

The rate at which magnetic helicity on some large scale $l$ is dissipated is $\eta \langle {\bf J}\cdot{\bf B}\rangle_l$, or in other words proportional to the current helicity averaged over the same scale. The large-scale average of the current helicity is the sum of contributions from smaller scales. We can find the expected distribution of current helicity over scales by considering the transfer rate between scales, $2{\bf B}_l\cdot({\bf v\times b})_{kl}$ where a positive sign represents the transfer of magnetic helicity to large scale structure. In a stationary state, this goes to zero as the resistivity goes to zero. The usual expansion of $({\bf v\times b})_l$ to first order in the eddy correlation time (\citep{Pouquet76}) gives four terms: a contribution proportional to large-scale average magnetic helicity embedded in small-scale eddies, $h_{kl}$, a similar contribution due to the kinetic helicity, but with the opposite sign, a term proportional to the gradient of the large scale field - usually written as a turbulent diffusivity, and a drift term of the form ${\bf V}_{eff}{\bf \times B}_l$. Obviously, the last term is orthogonal to ${\bf B}_l$ is doe snot contribute to the transfer of magnetic helicity between scales. The first term leads to the transfer of $h_{kl}$ to larger scales at a rate comparable to $\tau_k^{-1}$, the small scale eddy turnover rate\citep{Vishniac_Cho01}). The second term depends on kinetic helicity, which is not a conserved quantity. Naively one would expect it to be embedded in small scales only to the extent that the symmetry breaking of the large-scale environment biases the small-scale turbulence. However, the current helicity embedded on small scales induces a kinetic helicity, whose effect is to drive the second term to cancel a fraction of the first term (\citep{Vishniac_Shapovalov14}). Finally, the third term contributes an effect which is approximate $-\langle (v_iv_j+b_ib_j\rangle_l\epsilon_{ist}B_{ls}\partial_j B_{lt}\tau_k$.
(Here we have neglected a contribution due to pressure in the force equation, which somewhat reduces the contribution from the correlated small-scale magnetic fields). This term is off order $-\langle v^2\rangle_L\tau_k$ times the large-scale contribution to the current helicity. This is the term that has to balance the first two terms. Since the small-scale contribution to the current helicity is $k^2h_{kl}$, the ratio of the small-scale current helicity to the large-scale current helicity is $~(kv_k\tau_k)^2$, i.e. of order unity. The current helicity, and consequently the ohmic dissipation of magnetic helicity, is distributed across the full range of the inertial cascade with a one-dimensional spectral index of $-1$.
 This in turn implies that the dissipation of magnetic helicity will be enhanced by some logarithmic factor above the ohmic dissipation rate on the large eddy scale. Simulations that use artificial forms of resistivity that depend on higher powers of the wavenumber will damp magnetic helicity at slower rates because of the concentration of dissipation at the smallest eddy scales and because those eddies will be smaller. However, the reduced range of scales subject to dissipation will tend to accumulate extra power, and presumably magnetic helicity, due to the bottleneck effect which will work in the opposite direction.

 In a turbulent cascade the energy dissipation rate is the energy density times the large-scale eddy turnover rate, i.e. the characteristic time scale is just $L/v_L$.  We have shown that the corresponding time for the dissipation of magnetic helicity is  $~L^2/\eta$ (divided by some logarithmic measure of the dynamic range of the turbulent cascade). Dividing the former by the latter gives us a measure of how well magnetic helicity is conserved on dynamic times scales. This is inversely proportional to the magnetic Reynolds number $V_L L/\eta$, which in most cases means that magnetic helicity is conserved to a level that cannot be adequately reproduced relying only on numerical precision.

 There is one final point worth stressing about the dissipation of magnetic helicity at different scales within the turbulent cascade.  We can derive a maximum dissipation rate without making an explicit calculation of the transfer of magnetic helicity between scales.  On dimensional grounds, the maximum current helicity at a given wave number will be $\sim k b_k^2$. However, highly magnetized turbulence will be highly anisotropic and the current helicity depends on symmetry breaking in all three directions. Consequently, the actual limit is $k_\| b_k^2$. If we multiply this by the Alfven speed we get the turbulent energy density times the Alfven frequency for typical eddies with wave number $~k$. Invoking the condition of critical balance, $\tau_{nonlinear}^{-1}\sim \omega_A$ which is a generic feature of MHD turbulence models, we see that the maximum average current helicity embedded in eddies of wave number $k$ is proportional to the energy cascade rate at those scales. From the conservation of energy and locality of nonlinear transfer of energy in phase space, this shows that the maximum dissipation rate of magnetic helicity is constant within the turbulent cascade. In other words, the rate we derived in the previous paragraph, using a detailed but approximate analysis of the transfer of magnetic helicity in phase space, is roughly the same as the maximum rate we could expect in any model of MHD turbulence.

The problem we wish to address in this paper is that the distribution of magnetic helicity will diverge from the distribution given by evolving eqs. (\ref{Hflux} and (\ref{Hevolve}) with $\eta=0$, due to small numerical errors induced by discreteness effects, and by the necessity of using $\eta>0$ (or its equivalent) to promote numerical stability. This leads to a magnetic field ${\bf B}_{calc}$ which differs from the ideal result on small scales and whose associated magnetic helicity is not conserved \citep{Bodo+17}. Of course, we can always redefine our goal as a simulation with significant resistivity, but this fails to accurately model the dynamics of realistic astrophysical systems. We can define our error, i.e. the amount by which we have failed to treat the magnetic helicity as a locally conserved quantity by calculating $\nabla\cdot {\bf J}_H$ at every time step and advancing $H$ accordingly. Tracking magnetic helicity in a simulation requires calculating $\nabla\cdot {\bf J}_H$, which implies a knowledge of $\bf A$ and $\phi$ everywhere.
If the underlying code is pseudospectral or is at least based on a regular grid, then this is straightforward \citep{Squire&Bhattacharjee16}. Otherwise, the code has to be supplemented with Green's functions-based routines that will calculate these quantities \citep{Yousef+08, Singh&Jingade15}.

Here we propose a rigorous procedure that will bring the magnetic helicity back into alignment with its expected value while minimizing the size of the correction.
We could write this as a straightforward Gaussian minimization problem by defining an action dependent on the mean square correction to the field, subject to constraints that ensure that the corrected field will have the correct magnetic helicity distribution and will satisfy the usual constraints on the magnetic field. This is
\begin{equation}
S\equiv \int [(1/2)(\delta{\bf B})^2+\lambda({\bf x}) (H-{\bf A\cdot B})+\mu({\bf x})\nabla\cdot{\bf B}] dV,
\end{equation}
where $H$ is the magnetic helicity distribution that we expect and $A$ and $B$ are the corrected values of the vector potential and magnetic field. The last term in the integral enforces the usual zero divergence condition on the magnetic field. Without this term the corrected ${\bf B}$ would no longer satisfy Maxwell's equations. It is not necessarily important to obtain the correct distribution of $H$ on all scales. Just correcting it on scales comparable to, and larger than, the large-scale eddy size would improve the accuracy of dynamo simulations. Random fluctuations in the magnetic helicity flux produce random fluctuations in the amplitude of magnetic helicity modes with scales comparable to the large-scale eddies or smaller. The large wavelength tail of these fluctuations may be interesting, but small errors in the amplitude of eddy scale helicity fluctuations are unlikely to have any dynamical significance. Consequently, we are primarily concerned with correcting the distribution of magnetic helicity at wave numbers significantly smaller than the inverse of a large eddy scale.

\label{fig:Map_B_A}

Minimizing the change to the magnetic field is not the only choice. It's more convenient and more physically reasonable to minimize the change to the vector potential. The same change in the vector potential at small scales produces a larger change in B, so the net effect is that this choice puts more of the adjustment in the magnetic field at small scales. (This is not equivalent to correcting the magnetic helicity distribution at small scales. The magnetic field on small scales contributes to the large-scale distribution of magnetic helicity.) This is a better choice since we're trying to counteract the effects of inaccuracies at small scales. In addition, an algorithm that compensates for lost magnetic helicity by altering the large-scale magnetic field runs the risk of driving a large-scale dynamo by fiat. Conversely, fixing errors in the magnetic helicity distribution by making correlated changes on small scales implies that the bulk of the corrections will occur on scales within the inertial range of the turbulent cascade.  On these scales magnetic helicity is passed between scales at eddy turnover rates and any excess energy is quickly dissipated. If we weight the algorithm by $|\delta {\bf A}|^2$ then the action we choose to minimize is
\begin{equation}
 S\equiv \int [(1/2)(\delta{\bf A})^2+\lambda({\bf x}) (H-{\bf A\cdot B})-\mu({\bf x}) \nabla\cdot{\bf A}] dV.
 \end{equation}
The second Lagrangian constraint now fixes the gauge of ${\bf A}$, since the usual divergence-free condition for ${\bf B}$ is automatically satisfied. Integrating this by parts involves a surface term
\begin{equation}
   \Delta S=\int\left[\lambda{\bf A\times\delta A-\mu \delta A}\right]\cdot dS. 
\end{equation}In a finite computational volume this gives us a pair of boundary conditions, that $\lambda$ and $\mu$ vanish nearly everywhere on the boundary, the only exception being places where $\bf A$ is normal to the surface (for $\lambda$). This guarantees that there is no nontrivial solution for $\lambda$ unless the calculated distribution of magnetic helicity is nonzero somewhere. Here we will assume a periodic box and ignore this constraint. We obtain 
\begin{equation}
\delta{\bf A}=2\lambda {\bf B}+\nabla(\lambda)\times{\bf A}-\nabla\mu,
\label{newA}
\end{equation}
and
\begin{equation}
H={\bf A\cdot B}.
\end{equation}

The second equation is just a restatement of our goal, that the field quantities should give the correct distribution of magnetic helicity. The first equation is not as simple as it appears, since the fields on the RHS are the corrected fields, not the calculated ones. The last term on the RHS arises from the Coulomb gauge and leads to
\begin{equation}
\nabla^2\mu=\nabla\lambda\cdot{\bf B}=\nabla\cdot(\lambda{\bf B}).
\end{equation}


In order to make this problem more tractable, and to reduce the numerical cost of implementing a correction scheme, we will assume that $\lambda$ is small and that we can therefore ignore the difference between the corrected and computed fields on the RHS of the first equation. This should be reasonable if the correction scheme is applied often enough. If we further assume that the correction to the magnetic helicity can be approximated to the same order, i.e. linear in $\lambda$, then we have a second-order differential equation for $\lambda(x)$ in the computational box. With some work, this can be shown to be

\begin{align}
\Delta H\equiv& H-({\bf A}_{calc}\cdot{\bf B}_{calc})\\
\approx& \delta{\bf A}\cdot {\bf B} + {\bf A}\cdot (\nabla \times \delta{\bf A})\nonumber \\ \nonumber
=&2\lambda B^2+{\bf B}\cdot(\nabla\lambda\times {\bf A})-{\bf B}\cdot\nabla\mu+2\lambda{\bf J\cdot A}
\\
+&2{\bf A}\cdot(\nabla\lambda\times{\bf B})
+A_iA_j\partial_i\partial_j\lambda-{1\over2}
\nabla\lambda\cdot\nabla(A^2)-A^2\nabla^2\lambda
\nonumber\\
=&2\lambda(B^2+{\bf A\cdot J})-\nabla\lambda\cdot({\bf A\times\nabla\times A})-{1\over2}\nabla\lambda\cdot\nabla(A^2)
\nonumber\\
&+(A_iA_j-A^2\delta_{ij})\partial_i\partial_j\lambda
\nonumber
\\
= &{2(B^{2} + {\bf A\cdot J})\lambda +\partial_i\left( (A_iA_j -A^2\delta_{ij})\partial_j\lambda\right)-{\bf B}\cdot\nabla\mu}\nonumber
\label{lambda}
\end{align}
where we employ implicit summation over repeated indices.
The last term can be written as 
\begin{equation}
{\bf B}\cdot\nabla\mu={\bf B}\cdot\nabla\nabla^{-2}\nabla\cdot(\lambda{\bf B}).
\label{gauge}
\end{equation}
The operator $\nabla\nabla^{-2}\nabla$ is just $\hat k\hat k$ in Fourier space. In real space, it is an integral operator acting on everything to its right and can be written
\begin{align}
    \nabla \nabla^{-2}\nabla\cdot ({\bf F})=&{{\bf F}\over3}\\
    +&\int {d^3{\bf r'}\over 4\pi r'^5}\left(3{\bf r'r'\cdot F(r+r')}-r'^2
    {\bf F(r+r')}\right)\nonumber,
\end{align}
where ${\bf F}$ is any differentiable vector field.
As before, we are assuming that boundary conditions can be ignored. To linear order, we obtain $\lambda$ from 
\begin{eqnarray}
\Delta H=2(B^2+{\bf A\cdot J})\lambda+&\partial_i\left( (A_iA_j -A^2\delta_{ij})\partial_j\lambda\right)\nonumber\\&-{\bf B}\cdot\nabla \nabla^{-2} \nabla(\lambda{\bf B}),
\label{lambda1}
\end{eqnarray}
and use this in equation (\ref{newA}).


We need a general procedure for solving eq.(\ref{lambda1}).
We can do this through iteration. If we split two of the coefficients on the RHS into varying and constant pieces defined by
\begin{equation}
C_0\equiv \langle B^2+{\bf J}\cdot{\bf A}\rangle=2\langle B^2\rangle,
\end{equation}
\begin{equation}
C_1\equiv B^2+{\bf J}\cdot{\bf A}-C_0,
\end{equation}
\begin{equation}
D_{0ij}\equiv\langle \delta_{ij}A^2-A_iA_j\rangle,
\end{equation}
and
\begin{equation}
D_{1ij}\equiv \delta_{ij}A^2-A_iA_j-D_{0ij},
\end{equation}
then we can rewrite equation (\ref{lambda1}) as
\begin{eqnarray}
\Delta H-2C_1\lambda+\partial_iD_{1ij}\partial_j\lambda+&{\bf B}\cdot\nabla\left(\nabla^{-2}\nabla\cdot(\lambda{\bf B})\right)=\nonumber\\ &2C_0\lambda-\partial_iD_{0ij}\partial_j\lambda.
\label{itera}
\end{eqnarray}
The fourth term on the LHS of eq.(\ref{itera}) resists any simple reduction since the operator ${\bf B}\nabla^{-2}{\bf B}$ always depends on the wavenumber of $\lambda({\bf k})$. In Fourier space, the RHS of this equation is a simple multiplication by the positive definite coefficient which is always greater than $2C_0$.

\begin{eqnarray}
\lambda_{n+1}({\bf k})=&{1\over 2C_0+k_ik_jD_{0ij}}\Bigl[\Delta H({\bf k}) + \Bigl(-2C_1\lambda_n+\partial_iD_{1ij}\partial_j\lambda_n +\nonumber\\ 
&{\bf B}\cdot\nabla\left(\nabla^{-2}\nabla\cdot(\lambda_n{\bf B})\right)\Bigr)({\bf k})\Bigr].
\label{itera1}
\end{eqnarray}

The choice of $\lambda_0$ is not dictated by this method. One possiblity is to use $\langle \Delta H\rangle/2C_0$ as $\lambda_0$ so that
\begin{equation}
    \lambda_1={1\over 2C_0+k_ik_jD_{0ij}}\left[\Delta H({\bf k})-C_1\langle\Delta H\rangle/C_0\right].
\end{equation}
Alternatively, we could choose
\begin{equation}
    \lambda_0={\Delta H\over 2(C_0+C_1)}.
\end{equation}
\\

The former captures the direct response to a helicity error at a wavenumber ${\bf k}$ and a small part of the indirect response through the interaction with the spatial variation in the coefficient of $\lambda$. The latter captures the nonlinear part a bit better but misses the dependence on gradients of $\lambda$.

We can get a sense of how important the various contributions to $\lambda$ are by considering two simple cases. Note that due to random fluctuations in the magnetic helicity flux, we expect random fluctuations in the amplitude of modes with scales comparable to the large-scale eddies or smaller. Small errors in the amplitude of these fluctuations are unlikely to have any dynamic significance.  Consequently, we are primarily concerned with correcting the distribution of magnetic helicity at wave numbers significantly smaller than the inverse of a large eddy scale. For simplicity, we will assume that $\Delta H({\bf k})$ is nonzero only for these modes larger than the large-scale eddy modes. We expect the magnetic energy density will be largest at the large eddy scale, and the rms vector potential will be either largest at that scale or the large scale magnetic field domain scale.  In the former case, we have $C_0\sim B_T^2\sim D_{0ij}k_T^2$, where $k_T$ is the wave number of the large-scale eddies. The fluctuating coefficients $C_1$ and $D_{1ij}$ will be functions of scale, but at the large eddy scale, both are comparable to their nonfluctuating pieces. We see from eq.(\ref{itera}) that the contribution to $\lambda$ from the large eddy scale will be comparable to that from the scales where $\Delta H$ is nonzero. On smaller scales, $l<<l_T$ dominant term in $C_1$ will be ${\bf A\cdot J}$, which will be of order $A_TA_l/l^2\propto l^{-2/3}$ for the Goldreich-Sridhar model of turbulence. Consequently the contribution to $\lambda$ from scales in the inertial cascade will decrease as $l^{4/3}$

The magnetic domain scale vector potential can be larger than $A_T$ even in the early stages of the large-scale dynamo. In this case, the previous argument goes through in a slightly different form. Now the contribution to $\lambda$ from the turbulent scale is reduced by a factor of $A_T/A_0$ relative to the scale where $\Delta H$ is nonzero. As before the contribution to $\lambda$ falls off as $l^{4/3}$ on smaller scales.

We conclude that solutions for $\lambda$ need to be calculated to scales somewhat smaller than the large eddy scale in order to recover the desired distribution of magnetic helicity on magnetic domain scales.

 After solving for $\lambda$ to the required accuracy we correct the magnetic potential:
\begin{equation}
\delta{\bf A}={\bf\nabla\times}(\lambda{\bf A})+\lambda{\bf B}-\nabla\nabla^{-2}{\bf\nabla\cdot}(\lambda{\bf B}).
\label{answer}
\end{equation}
In phase space the last term does not have to be evaluated explicitly, rather the sum of the first two terms can simply be projected onto the plane perpendicular to ${\bf k}$.
If the initial error in the magnetic helicity distribution is large then it may be necessary to iterate to allow for the nonlinear term $(\bf\nabla\times \delta A)\cdot \delta A$.

If we neglect the gradient of $\lambda$ and consider the correction to magnetic helicity, we find that
\begin{equation}
    \Delta H\approx {\bf B\cdot (2\lambda B)}+{\bf A\cdot(\nabla\times 2\lambda B)}=2\lambda(B^2+{\bf A\cdot J}).
\end{equation}
The reconstituted magnetic helicity is not deposited strictly according to the energy density in the turbulent cascade, but on average it works out to the same thing as a function of scale. This implies that the power spectrum of the deposited magnetic helicity is much shallower than we would expect to find in a turbulent cascade. This is a consequence of 
eq. (\ref{answer}). There is some minimum scale, or maximum wavenumber, where the correction to the power spectrum is of order unity, $k_{max}\lambda\sim 1$. In order to avoid highly unphysical effects on the small-scale turbulent spectrum we require that the dissipation scale wavenumber satisfy
\begin{equation}
   1> k_{diss}\lambda \sim {k_{diss}\Delta H\over 4\langle B^2\rangle}.
\end{equation}
We can use this to set a limit on the cadence of corrections since we need 
\begin{equation}
\Delta H <4\langle B^2\rangle l_{diss}.
\label{deltaH_diss}
\end{equation}

\section{Testing the Algorithm}

It is not immediately obvious that the procedure described in eq. (\ref{itera}) will converge efficiently, or at all. We have performed a simple test using a time slice from a simulation of MHD turbulence in the turbulence database of The Johns Hopkins University ( \citep{2008JTurb...9...31L,2013Natur.497..466E}. In this simulation, the turbulence is driven at wavenumbers of 2 and contains an inertial range extending roughly from wavenumbers of $10^1$ to $10^2$. Rather than tracking the magnetic helicity over time and using our algorithm to correct to a predicted magnetic helicity distribution we deliberately distort the magnetic field and treat the new magnetic helicity distribution as an error that the algorithm will correct. To do this we take each Fourier mode in our randomly chosen time slice and change the phase randomly, with an rms value of $0.5$. For computational convenience, we degrade the resolution by a factor of $2$ to $512^3$. This eliminates about half of the strongly damped regime but leaves the inertial cascade untouched.  
We can check the scale distribution of helicity dissipation by plotting the one-dimensional spectrum of the current helicity or
${\bf J}(-{\bf k})\cdot {\bf B}({\bf k)}k^2$.  We see a bump at small wavenumbers and rough agreement with our predicted slope from wave numbers $|{\bf k}|\sim 20$ to well into the damped regime. We note that this implies that some of the dissipation is occurring well within the damped regime where the magnetic spectrum is sharply declining.

One complication is that the gauge term given in eq. (\ref{gauge}) is nonlocal and potentially expensive to evaluate. In Fourier space, the central operator has a simple form, but calculating nonlinear terms in real space means that we need to perform two fast Fourier transforms every time we evaluate this expression. Instead, we use a local estimate of this term following the method described by \citet{Wagneretal2017}.

\begin{figure*}
\includegraphics[width=0.49\linewidth]{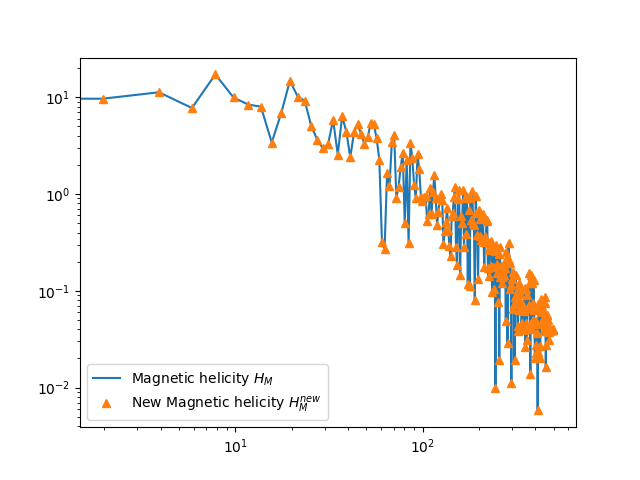}
\includegraphics[width=0.49\linewidth]{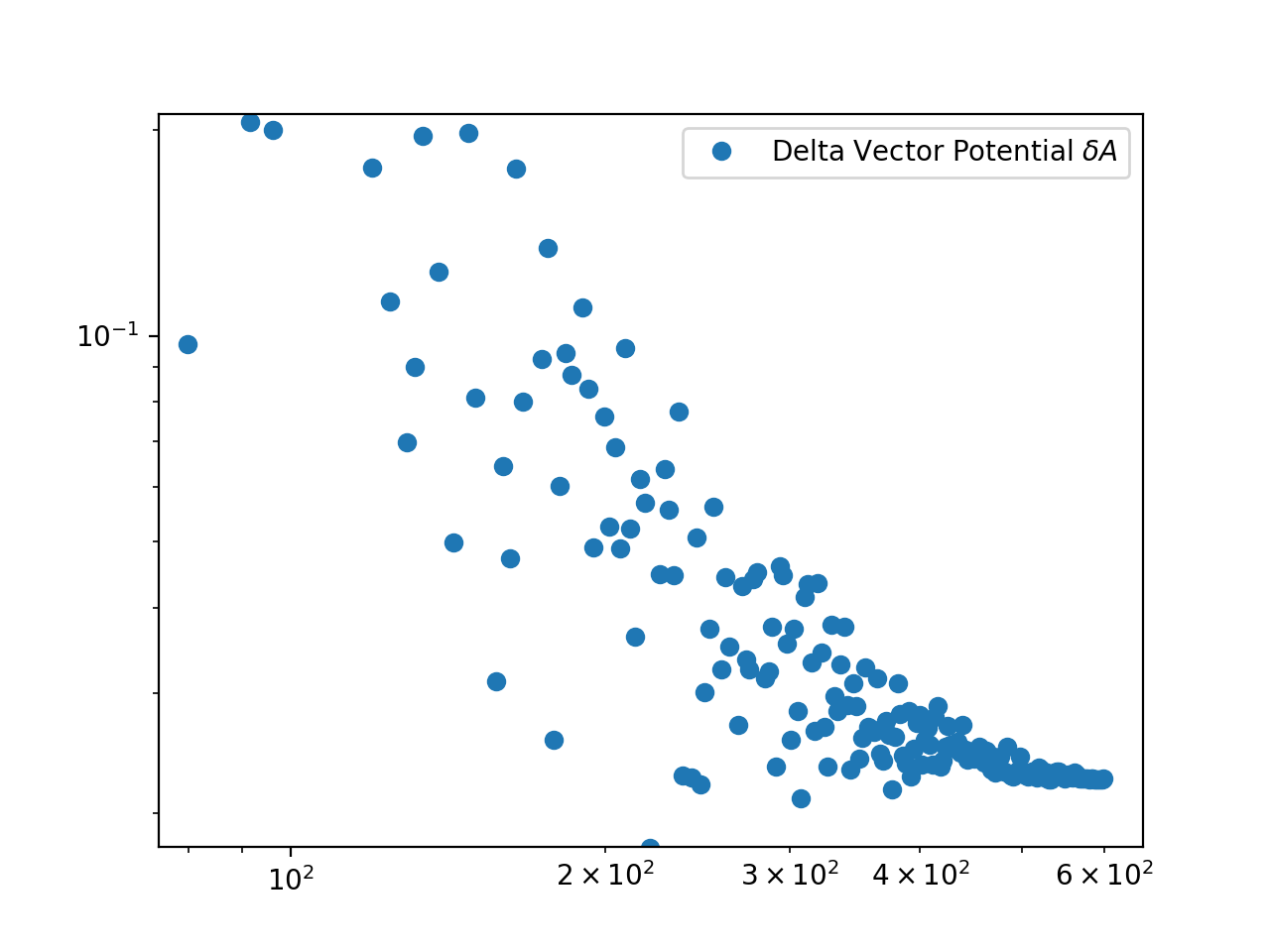}
\caption{Left: Log scale of the power spectrum of the original magnetic helicity and final magnetic helicity after modification by eq.(\ref{lambda1}) as a function of modes. Right: Log scale of the power spectrum of the delta of original vector potential A and the final vector potential after modified by eq.(\ref{answer}) as a function of modes}
\label{fig:PS_Helicity_VecP}
\end{figure*}

Applying the algorithm to the altered map and using the original helicity distribution as the desired goal, we obtain convergence to the limit of numerical accuracy for $\lambda$ at all scales in between 6 and 30-time steps. In fig. (\ref{fig:PS_Helicity_VecP}) we show a comparison between the initial distribution of helicity and the magnetic potential as a function of wavenumber and the final values after $\lambda$ converged via the repeated application of eq. (\ref{itera1}). These plots show amplitudes, and not phase information, but since phase changes lead to amplitude changes in the helicity, this is an adequate proxy for the ability of the algorithm to recover the magnetic helicity. From the second panel, we see that we recover the original distribution of magnetic helicity without introducing significant errors in the vector potential.

\begin{figure}
\includegraphics[width=\linewidth]{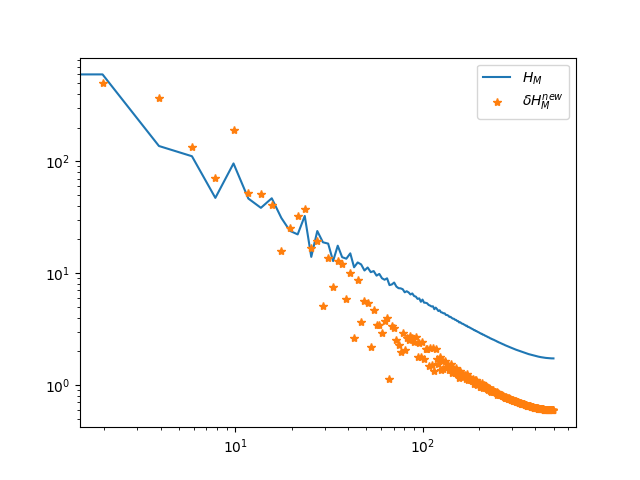}
\caption{The magnetic helicity H (solid line) from the simulation data compared to the change produced by a smaller phase shift (rms $0.05$) of the Fourier modes. In this case the change in $\Delta H$ (dotted point) calculated by eq.(\ref{itera}) is significantly smaller than H.}
\label{fig:dHsmall}
\end{figure}

Finally, we performed a test with a smaller shift in the phase of the modes (an rms phase shift of $0.05$).  In this case, the change in the magnetic helicity, and subsequent corrections, were small at higher wavenumbers, although still substantial at large wavelengths. This is illustrated in fig. (\ref{fig:dHsmall}).

\section{DISCUSSION and CONCLUSIONS}

Magnetic helicity is an important topological quantity that constrains large-scale relaxation of magnetized systems and plays an important role in large-scale dynamos.  Since the rate at which it is dissipated in a turbulent cascade is smaller than the large-scale turnover rate (or the energy dissipation rate) requiring a simulation to treat it as a locally conserved quantity is roughly as important as requiring mass conservation.
In this study, we have constructed an algorithm for correcting errors in the magnetic helicity distribution in numerical simulations. For the purposes of this work, we define an error as any deviation between the evolution of the simulated system and one in which magnetic helicity is a locally conserved quantity, as it typically is in the systems being modeled.  Numerical simulations that include turbulence frequently show a loss in global magnetic helicity. There is no published information on the ability of current algorithms to track magnetic helicity correctly on smaller scales but it is reasonable to suppose that they do significantly worse at this more difficult task. While there is no unique way to restore lost information, the algorithm explored here has some attractive features, including the ability to embed the lost helicity to roughly the same scales where numerical errors eliminated it. Adopting this algorithm has two associated costs. The most obvious is the computational costs involved with the algorithm itself. However, this method cannot be applied without tracking local magnetic helicity in a simulation, i.e. without the use of equations (\ref{Hflux}) and (\ref{Hevolve}) with $\eta=0$. The latter implies a fractional increase in the computational costs of every timestep. The former is hard to predict without knowing the necessary cadence of corrections but is probably less significant. We note also that in the event that the resistivity is large enough that its role in the system evolution is not entirely swamped by numerical noise that this algorithm will still function properly as long the resistivity is included in eqs. ({\ref{Hflux}}) and ({\ref{Hevolve}}).

We have examined the cleaning algorithm for magnetic helicity in the JHU $1024^3$ MHD turbulence simulation. The algorithm restored the large-scale distribution of magnetic helicity after it was distorted by randomly altering the phases of Fourier components of the magnetic field. The iterative procedure for constructing the solution converged, despite a lack of formal proof of convergence.  As expected the algorithm does increase the amount of very small-scale power in the turbulent cascade, a result which needs to be considered in setting the cadence of correction in a simulation.  We also note that while magnetic helicity is robustly conserved with any gauge choice our work is aimed specifically at enforcing conservation in the Coulomb gauge. We justify this choice by pointing to the close correlation between the magnetic helicity in this gauge and the current helicity, which appears in analytic theories of the large-scale dynamo (and is not gauge dependent).

One complication is that we have not formally demonstrated that our procedure for deriving $\lambda$ will always converge. We have shown, as a practical matter, that it converges in our test case, a simulation of homogeneous isotropic magnetized turbulence. This is a sufficiently robust test that it is plausible that it will always converge. However, we note that the fluctuating parts of the coefficients in the second-order equation for $\lambda$, eq. (\ref{lambda1}), will be roughly as large as the constant parts. Moreover, the gauge constraint term will typically be of order $\lambda B^2$, i.e. the same order as the first RHS term. We have already seen that in real space this term is the sum of $\lambda B^2/3$ plus a local quadrupole contribution which we can reasonably expect to be smaller. We can compare it to the first RHS term, whose average value is $4B^2\lambda$. We see that the 
 gauge constraint term will be smaller by an order of magnitude than the other terms on the RHS.
 
 The difficulty of enforcing strict conservation of magnetic helicity in MHD simulations is a major obstacle in performing realistic simulations of large-scale dynamo processes in astrophysical objects. This is particularly true for magnetorotational instability, where the large-scale eddy size can vary dramatically over time.
 More generally, magnetic helicity conservation may play a key role in the shear-current effect \citep{Squire&Bhattacharjee16}. Moreover, for small-scale MHD fluctuations, this effect may define the off-diagonal turbulent diffusivity $\eta_{xy}$ \citep{Singh&Jingade15, Squire&Bhattacharjee16, Singh&Jingade15, Naveen&Singh21, Dewar+20}. As a result, this could be the crucial bound of for MHD instabilities.
 Our results show that a viable correction scheme that exists and can be used to produce a realistic magnetic helicity distribution. 
 In future work, we will include this cleaning algorithm in a stratified periodic shearing box simulation (see, for example, \citet{Davis+10}) to follow the flow of magnetic helicity and its effect on the magnetorotational instability-driven dynamo.

\section{Acknowledgements}\label{Acknowledgements}
This work makes use of the Johns Hopkins Turbulence Database.
The authors wish to recognize and acknowledge helpful discussions with Amir Jafari and Greg Eyink. ETV thanks the AAS for supporting his research. 
YZ thanks Mor Rozner for stimulating discussions. YZ thanks the CHE Israelis Excellence Fellowship for Postdoctoral for supporting his research. Both authors wish to thank the referee, whose numerous questions led us to add detailed explanations of why magnetic helicity conservation is an important goal.  We believe this has led to a paper that is much more accessible to readers.

\bibliography{references}

\begin{thebibliography}{}
\expandafter\ifx\csname natexlab\endcsname\relax\def\natexlab#1{#1}\fi
\providecommand{\url}[1]{\href{#1}{#1}}
\providecommand{\dodoi}[1]{doi:~\href{http://doi.org/#1}{\nolinkurl{#1}}}
\providecommand{\doeprint}[1]{\href{http://ascl.net/#1}{\nolinkurl{http://ascl.net/#1}}}
\providecommand{\doarXiv}[1]{\href{https://arxiv.org/abs/#1}{\nolinkurl{https://arxiv.org/abs/#1}}}

\bibitem[{{Aluie}(2017)}]{Aluie2017}
{Aluie}, H. 2017, New Journal of Physics, 19, 025008,
  \dodoi{10.1088/1367-2630/aa5d2f}

\bibitem[{{Bodo} {et~al.}(2017){Bodo}, {Cattaneo}, {Mignone}, \&
  {Rossi}}]{Bodo+17}
{Bodo}, G., {Cattaneo}, F., {Mignone}, A., \& {Rossi}, P. 2017, \apj, 843, 86,
  \dodoi{10.3847/1538-4357/aa7680}

\bibitem[{{Brandenburg} {et~al.}(2009){Brandenburg}, {Candelaresi}, \&
  {Chatterjee}}]{brandenburg09a}
{Brandenburg}, A., {Candelaresi}, S., \& {Chatterjee}, P. 2009, \mnras, 398,
  1414, \dodoi{10.1111/j.1365-2966.2009.15188.x}

\bibitem[{{Brandenburg} \& {Subramanian}(2005)}]{Brandenburg+05}
{Brandenburg}, A., \& {Subramanian}, K. 2005, \physrep, 417, 1,
  \dodoi{10.1016/j.physrep.2005.06.005}

\bibitem[{{Candelaresi} {et~al.}(2011){Candelaresi}, {Hubbard}, {Brandenburg},
  \& {Mitra}}]{Candelaresi+11}
{Candelaresi}, S., {Hubbard}, A., {Brandenburg}, A., \& {Mitra}, D. 2011,
  Physics of Plasmas, 18, 012903, \dodoi{10.1063/1.3533656}

\bibitem[{{Davis} {et~al.}(2010){Davis}, {Stone}, \& {Pessah}}]{Davis+10}
{Davis}, S.~W., {Stone}, J.~M., \& {Pessah}, M.~E. 2010, \apj, 713, 52,
  \dodoi{10.1088/0004-637X/713/1/52}

\bibitem[{{Del Sordo} {et~al.}(2013){Del Sordo}, {Guerrero}, \&
  {Brandenburg}}]{DelSordo+13}
{Del Sordo}, F., {Guerrero}, G., \& {Brandenburg}, A. 2013, \mnras, 429, 1686,
  \dodoi{10.1093/mnras/sts398}

\bibitem[{{Dewar} {et~al.}(2020){Dewar}, {Burby}, {Qu}, {Sato}, \&
  {Hole}}]{Dewar+20}
{Dewar}, R.~L., {Burby}, J.~W., {Qu}, Z.~S., {Sato}, N., \& {Hole}, M.~J. 2020,
  Physics of Plasmas, 27, 062504, \dodoi{10.1063/5.0005740}

\bibitem[{{Eyink} {et~al.}(2013{\natexlab{a}}){Eyink}, {Vishniac}, {Lalescu},
  {Aluie}, {Kanov}, {B{\"u}rger}, {Burns}, {Meneveau}, \& {Szalay}}]{eyink13}
{Eyink}, G., {Vishniac}, E., {Lalescu}, C., {et~al.} 2013{\natexlab{a}}, \nat,
  497, 466, \dodoi{10.1038/nature12128}

\bibitem[{{Eyink} {et~al.}(2013{\natexlab{b}}){Eyink}, {Vishniac}, {Lalescu},
  {Aluie}, {Kanov}, {B{\"u}rger}, {Burns}, {Meneveau}, \&
  {Szalay}}]{2013Natur.497..466E}
---. 2013{\natexlab{b}}, \nat, 497, 466, \dodoi{10.1038/nature12128}

\bibitem[{{Gruzinov} \& {Diamond}(1996)}]{Gruzinov96}
{Gruzinov}, A.~V., \& {Diamond}, P.~H. 1996, Physics of Plasmas, 3, 1853,
  \dodoi{10.1063/1.871981}

\bibitem[{{Hubbard} \& {Brandenburg}(2011)}]{Alexander&Axel11}
{Hubbard}, A., \& {Brandenburg}, A. 2011, \apj, 727, 11,
  \dodoi{10.1088/0004-637X/727/1/11}

\bibitem[{{Ji} {et~al.}(1996){Ji}, {Prager}, {Almagri}, {Sarff}, {Yagi},
  {Hirano}, {Hattori}, \& {Toyama}}]{Ji96}
{Ji}, H., {Prager}, S.~C., {Almagri}, A.~F., {et~al.} 1996, Physics of Plasmas,
  3, 1935, \dodoi{10.1063/1.871989}

\bibitem[{{Jingade} \& {Singh}(2021)}]{Naveen&Singh21}
{Jingade}, N., \& {Singh}, N.~K. 2021, arXiv e-prints, arXiv:2103.12599.
\newblock \doarXiv{2103.12599}

\bibitem[{{Li} {et~al.}(2008){Li}, {Perlman}, {Wan}, {Yang}, {Meneveau},
  {Burns}, {Chen}, {Szalay}, \& {Eyink}}]{2008JTurb...9...31L}
{Li}, Y., {Perlman}, E., {Wan}, M., {et~al.} 2008, Journal of Turbulence, 9,
  N31, \dodoi{10.1080/14685240802376389}

\bibitem[{{Mitra} \& {Brandenburg}(2012)}]{Mitra_Brandenburg12}
{Mitra}, D., \& {Brandenburg}, A. 2012, \mnras, 420, 2170,
  \dodoi{10.1111/j.1365-2966.2011.20190.x}

\bibitem[{{Pouquet} {et~al.}(1976){Pouquet}, {Frisch}, \& {Leorat}}]{Pouquet76}
{Pouquet}, A., {Frisch}, U., \& {Leorat}, J. 1976, Journal of Fluid Mechanics,
  77, 321, \dodoi{10.1017/S0022112076002140}

\bibitem[{{Singh} \& {Jingade}(2015)}]{Singh&Jingade15}
{Singh}, N.~K., \& {Jingade}, N. 2015, \apj, 806, 118,
  \dodoi{10.1088/0004-637X/806/1/118}

\bibitem[{{Squire} \& {Bhattacharjee}(2016)}]{Squire&Bhattacharjee16}
{Squire}, J., \& {Bhattacharjee}, A. 2016, Journal of Plasma Physics, 82,
  535820201, \dodoi{10.1017/S0022377816000258}

\bibitem[{{Taylor}(1974)}]{Taylor74}
{Taylor}, J.~B. 1974, \prl, 33, 1139, \dodoi{10.1103/PhysRevLett.33.1139}

\bibitem[{{Taylor}(1986)}]{Taylor86}
---. 1986, Reviews of Modern Physics, 58, 741,
  \dodoi{10.1103/RevModPhys.58.741}

\bibitem[{{Vishniac} \& {Brandenburg}(1997)}]{Vishniac_Brandengurg97}
{Vishniac}, E.~T., \& {Brandenburg}, A. 1997, \apj, 475, 263,
  \dodoi{10.1086/303504}

\bibitem[{{Vishniac} \& {Cho}(2001)}]{Vishniac_Cho01}
{Vishniac}, E.~T., \& {Cho}, J. 2001, \apj, 550, 752, \dodoi{10.1086/319817}

\bibitem[{{Vishniac} \& {Shapovalov}(2014)}]{Vishniac_Shapovalov14}
{Vishniac}, E.~T., \& {Shapovalov}, D. 2014, \apj, 780, 144,
  \dodoi{10.1088/0004-637X/780/2/144}

\bibitem[{{Wagner} {et~al.}(2017){Wagner}, {Dangel}, {Cartarius}, {Main}, \&
  {Wunner}}]{Wagneretal2017}
{Wagner}, M., {Dangel}, F., {Cartarius}, H., {Main}, J., \& {Wunner}, G. 2017,
  Acta Polytecnica, 57, 470, \dodoi{https://doi.org/10.14311/AP.2017.57.0470}

\bibitem[{{Woltjer}(1958)}]{Woltjer58}
{Woltjer}, L. 1958, Proceedings of the National Academy of Science, 44, 489,
  \dodoi{10.1073/pnas.44.6.489}

\bibitem[{{Yamada}(1999)}]{Yamada99}
{Yamada}, M. 1999, Washington DC American Geophysical Union Geophysical
  Monograph Series, 111, 129, \dodoi{10.1029/GM111p0129}

\bibitem[{{Yousef} {et~al.}(2008){Yousef}, {Heinemann}, {Rincon},
  {Schekochihin}, {Kleeorin}, {Rogachevskii}, {Cowley}, \&
  {McWilliams}}]{Yousef+08}
{Yousef}, T.~A., {Heinemann}, T., {Rincon}, F., {et~al.} 2008, Astronomische
  Nachrichten, 329, 737, \dodoi{10.1002/asna.200811018}

\end{thebibliography}
\end{document}